# Kinematic body responses and perceived discomfort in a bumpy ride: Effects of sitting posture

Marko Cvetkovic, Raj Desai, Georgios Papaioannou, & Riender Happee

Faculty of Mechanical Engineering, Delft University of Technology, 2628 CD Delft, The Netherlands

## ABSTRACT

The present study investigates perceived comfort and whole-body vibration transmissibility in intensive repetitive pitch exposure representing a bumpy ride. Three sitting strategies (preferred, erect, and slouched) were evaluated for perceived body discomfort and body kinematic responses. Nine male and twelve female participants were seated in a moving-based driving simulator. The slouched posture significantly increased lateral and yaw body motion and induced more discomfort in the seat back area. After three repetitive exposures, participants anticipated the upcoming motion using more-effective postural control strategies to stabilize pelvis, trunk, and head in space.

## Introduction

Analysis of perceived discomfort is a complex and multifunctional problem that combines several crucial parameters. The variation of the road conditions, the whole-body vibration characteristics, the exposure, the postural adjustments, and the seat design have been extensively studied regarding their impact on occupant's perception of discomfort (Cvetkovic et al., 2021; Nawayseh et al., 2020). The road conditions (path, road profile and environment) can induce visual, vestibular, and muscular feedback, thereby provoking postural instability (i.e., more postural adjustments). The latter has a proven relation with local body discomfort, interfering with the center of mass' ability to sustain external perturbations (Mirakhorlo et al., 2022). To anticipate this, occupants' usually adopt different sitting strategies (re-posturing) while being driven (Cvetkovic et al., 2020). This greatly affects whole-body vibration transmissibility and perceived discomfort, but there is limited literature (Paddan & Griffin, 1994). Current works focus on either quantifying vibrational transmissibility (Nawayseh et al., 2020) neglecting the perceived comfort, or vice versa (Lecocq et al., 2022). To that end, this research quantifies the impact of adopted sitting strategies on kinematic body responses and perceived discomfort while exposed to intensive pitch vehicle motion through longitudinal perturbations. In particular, the research uncovers whether test subjects can reduce pelvis, trunk, and head angular rotation by adopting various sitting postures, and through anticipation.

## Methodology

The data collection included twenty-one test subjects, consisting of nine males and twelve females. The average age of the participants was 28.2 years ($\pm$ 4.7), with an average height of 170.6 cm ($\pm$ 7.5) and an average weight of 68.0 kg ($\pm$ 11.18). Prior to the experiment, all participants were informed of the experimental procedure and study via informed consent. Test subjects were offered a voucher for 20 EUR as compensation for participating in this study. The experimental design and procedure were in accordance with the Declaration of Helsinki. The procedure was recognized and approved by the Human Research Ethics Committee of the Delft University of Technology (HREC), under application number 962.

The participants were placed in a driving simulator (Figure 1a) with a standard passenger seat and exposed to intensive pitch motion representing a bumpy ride, and filtered to fit within the range of the motion platform. The perturbation was combined with modest fore-aft and lateral motion. More specifically, the input signal (Figure 1b) contained three bumps, with 0.55 m/s$^2$ rms power, separated by one second and lasting 15 seconds. Each test subject performed three experimental trials with different sitting strategies (i.e., preferred, erect, and slouched). The preferred posture aimed to achieve relaxed upper body muscles with an erect posture. In erect, the participants placed the buttocks at the most anterior position on the seat, pressing the abdomen and shoulders outwards, and keeping the back in an arched position. For the slouched posture, the participants moved their pelvis to the anterior seat pan position, flexing the lumbar spine towards a C-shape curvature. The upper arms stayed relaxed in the lap while keeping the chest and head straight. Table 1 lists the recorded posture for these 3 conditions.

Table 1. Flexion – Extension angles between the body segments (mean degrees ±standard deviation) and perceived comfort

|  | Postures | | |
|---|---|---|---|
|  | Erect | Preferred | Slouched |
| Head – Thorax | 13.7 ±10.7 | 13.0 ±8.7 | 23.1 ±11.1 |
| Pelvis – Thorax | 30.5 ±9.0 | 36.9 ±8.6 | 42.0 ±9.7 |

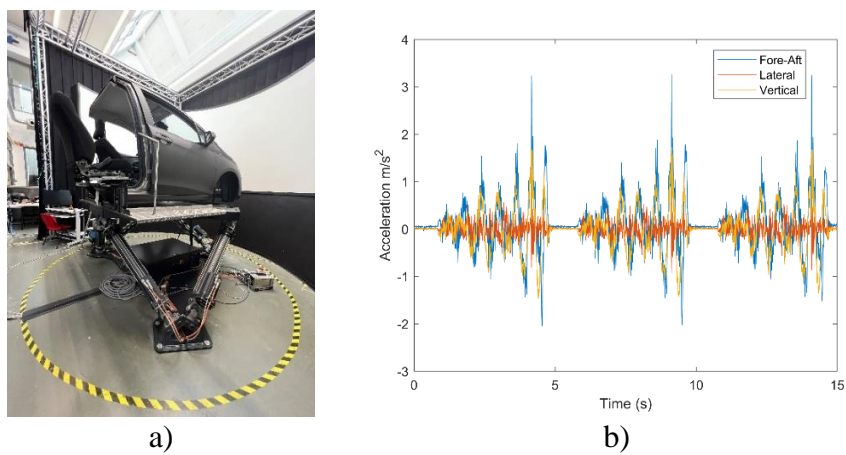

Figure 1. a) Delft Advanced Vehicle Simulator; b) recorded platform signal

Perceived motion discomfort was assessed using an 8-item questionnaire after each posture condition. The first item assessed motion sickness using the misery scale (MISC). The 2$^{nd}$ to 8$^{th}$ item addressed perceived discomfort in different body areas.

Kinematic 3D full body responses and adopted sitting strategies were captured using a motion-capturing system (MTW Awinda, Xsens Technologies, Enschede, The Netherlands). For the purpose of this work, the root mean square of the body segments (head, trunk, and pelvis) rotational kinematic responses (roll, pitch, and yaw) were calculated to estimate the effect of the adopted sitting posture (i.e., preferred, erect, and slouched) on the different body segments. Thus, we could quantify the variation in acceleration between the three different sitting strategies and determine whether the participants anticipated the bump after the repetitive exposure, by reducing the kinematic body responses.

Two-Way Repeated Measurements Analysis of variance (ANOVA) was conducted including three within-subjects factors (three postures, three body segments, and three bumps) to uncover the differences within dependent variables and its interactions. Additionally, an One-Way Repeated Measures ANOVA was performed to reveal differences between three bumps (B1, B2 and B3) for specific posture and body segment. The same method was performed to expose differences between perceived body discomfort and sitting postures. If criteria did not meet the Mauchly's Test of Sphericity (>0.05), Greenhouse-Geisser adjusted p-value was taken into consideration. A post hoc test, the Bonferroni adjustment, was utilized to reveal differences between collected variables and to determine between and within subject variation (B1-B2, B1-B3, and B2-B3).

**Results and Discussion**

Twenty-one participants completed the experiment, reporting acceptable motion comfort (Table 1) and negligible motion sickness (median, overall body comfort level, was 2.65 out of 10, with an Inter Quartile Range, IQR, of 0.36; median, MISC, was 0.38, IQR = 0.08). Due to insufficient and deviant kinematic data, two participants (males; mean age= 25.6 years; mean weight=77.5 kg; and mean body height = 177.5 cm) were excluded from further analysis.

As far as the perceived comfort is concerned, re-posturing from erect to slouched, resulted in more discomfort (p=0.008) in the lower back region (perceived comfort level/slouched = 3.14 ±1.06, and perceived comfort level/erect = 1.76 ±0.9). The postural effect on perceived discomfort was also reflected in reporting the backrest as the least comfortable in the slouched posture (p<0.001).

The kinematic body responses are displayed in Table 2. Regarding sitting postures, they substantially affected the pelvis, trunk, and head, roll and yaw responses (p=0.001 and p=0.043, respectively), while the pitch body rotational acceleration did not differ. Sitting erect and maintaining active posture resulted in lower roll and yaw rotational acceleration compared with more extensive pelvis-thorax angle – slouched posture (p-value for the roll = 0.004, yaw p=0.025). The erect sitting posture produced a stable center of mass and better posture alignment, resulting in lower body rotational acceleration versus slouched posture. As the participants were subjected to three perturbations, their body responses did not differ in the first two, but illustrated significant variations compared to the last perturbation. The repeated measurements ANOVA tests did not indicate significant differences between the first two consecutive bumps (B1-B2), except from pelvis pitch angular acceleration which differed (p=0.036) when participants adopted the preferred posture. According to the assessment of all body segments kinematic responses, the participants did anticipate the third motion by significantly reducing their roll, yaw, and pitch motion (p <0.001). The participants struggled more to stabilize their head (i.e., higher magnitude roll, pitch, and yaw responses than the other segments – Table 2) when exposed to the pitch platform motion. Meanwhile, more effective trunk-in-space stabilization was identified at the third bump (B3), where significantly lower roll, pitch, and yaw responses (Table 2) were measured for all sitting postures. Furthermore, the trunk illustrated more roll (p=0.008) and pitch (p=0.002) rotation compared to the pelvis, while the trunk was dominant over the head motion in yaw rotational acceleration (p=0.010). Similar to the trunk responses, the participants demonstrated improved stabilization of the pelvis during the third bump (Table 2).

Table 2. Mean rotational responses (± standard deviation) of body segment (head, trunk, and pelvis) in three sitting postures (erect, preferred, and slouched) during intensive-pitch platform motion. Analysis of significant differences and post hoc test of kinematic body responses in the three consecutive motions (B1, B2 and B3).

| Body segment | Posture | Bumps [deg/s$^2$] | | | | | | |
|---|---|---|---|---|---|---|---|---|
| | | B1 | B2 | B3 | p-diff | B1-B2 | B1-B3 | B2-B3 |
| *Roll kinematic responses* | | | | | | | | |
| Head | Erect | 182.4 ± 63.0 | 179.2 ± 66.4 | 168.2 ± 67.8 | *0.041$^b$* | n.s. | n.s. | *0.006* |
| | Preferred | 193.5 ± 105.6 | 188.7 ± 109.9 | 187.4 ± 96.2 | *0.853$^a$* | n.s. | n.s. | n.s. |
| | Slouched | 246.4 ± 245.7 | 227.9 ± 248.6 | 220.6 ± 202.3 | *0.232$^a$* | n.s. | n.s. | n.s. |
| Trunk | Erect | 233.8 ± 145.7 | 241.6 ± 147.7 | 192.3 ± 99.7 | *0.001$^a$* | n.s. | *0.005* | *0.007* |
| | Preferred | 281.9 ± 180.3 | 279.8 ± 191.0 | 221.4 ± 118.6 | *0.002$^b$* | n.s. | *0.004* | *0.014* |
| | Slouched | 276.2 ± 148.8 | 272.2 ± 139.7 | 244.5 ± 122.5 | *0.011$^b$* | n.s. | n.s. | *0.022* |
| Pelvis | Erect | 100.4 ± 35.6 | 99.1 ± 34.7 | 92.4 ± 37.5 | *0.018$^a$* | n.s. | *0.050* | n.s. |
| | Preferred | 141.6 ± 66.8 | 139.6 ± 68.2 | 128.8 ± 61.0 | *0.001$^a$* | n.s. | *0.011* | *0.004* |
| | Slouched | 240.4 ± 80.2 | 245.9 ± 81.7 | 228.3 ± 78.1 | *<0.001$^a$* | n.s. | n.s. | *0.004* |
| *Pitch kinematic responses* | | | | | | | | |
| Head | Erect | 455.9 ±132.1 | 463.2 ± 150.7 | 461.3 ± 155.6 | *0.078$^a$* | n.s. | n.s. | n.s. |
| | Preferred | 399.2 ± 97.5 | 423.1 ± 102.0 | 415.9 ± 75.1 | *0.135$^a$* | n.s. | n.s. | n.s. |
| | Slouched | 472.1 ± 356.7 | 475.9 ± 375.2 | 424.3 ± 224.4 | *0.213$^b$* | n.s. | n.s. | n.s. |
| Trunk | Erect | 339.3 ± 168.9 | 375.7 ± 199.9 | 303.4 ± 140.8 | *<0.001$^a$* | n.s. | *0.036* | *0.001* |
| | Preferred | 411.2 ± 245.2 | 411.3 ± 243.3 | 332.1 ± 176.1 | *<0.001$^a$* | n.s. | *0.002* | *0.001* |
| | Slouched | 376.8 ± 215.3 | 385.2 ± 222.4 | 313.3 ± 132.9 | *0.015$^b$* | n.s. | n.s. | *0.050* |
| Pelvis | Erect | 175.4 ± 37.4 | 170.5 ± 44.53 | 147.9 ± 23.6 | *0.004$^a$* | n.s. | *0.013* | *0.050* |
| | Preferred | 227.0 ± 107.0 | 236.0 ± 113.8 | 207.6 ± 100.0 | *<0.001$^a$* | *0.036* | *<0.001* | *<0.001* |
| | Slouched | 267.2 ± 109.1 | 272.6 ± 114.0 | 241.9 ± 110.1 | *0.001$^b$* | n.s. | *0.019* | *0.002* |
| *Yaw kinematic responses* | | | | | | | | |
| Head | Erect | 142.7 ± 56.2 | 118.2 ± 31.7 | 112.3 ± 28.4 | *0.056$^b$* | n.s. | n.s. | n.s. |
| | Preferred | 138.4 ± 78.7 | 131.4 ± 74.8 | 123.4 ± 64.1 | *0.018$^a$* | n.s. | *0.036* | n.s. |
| | Slouched | 199.7 ± 155.3 | 165.0 ± 104.9 | 156.0 ± 83.8 | *0.156$^b$* | n.s. | n.s. | n.s. |
| Trunk | Erect | 280.2 ± 162.7 | 282.0 ± 187.6 | 213.0 ± 118.5 | *<0.001$^a$* | n.s. | *<0.001* | *0.002* |
| | Preferred | 308.2 ± 227.4 | 317.1 ± 270.8 | 244.5 ± 186.0 | *<0.001$^a$* | n.s. | *0.001* | *0.008* |
| | Slouched | 246.3 ± 129.8 | 248.0 ± 132.2 | 210.2 ± 90.0 | *<0.001$^a$* | n.s. | *0.010* | *0.007* |
| Pelvis | Erect | 170.3 ± 66.0 | 168.0 ± 64.3 | 166.0 ± 72.7 | *0.477$^a$* | n.s. | n.s. | n.s. |
| | Preferred | 198.0 ± 66.2 | 196.2 ± 67.5 | 188.6 ± 65.6 | *0.002$^a$* | n.s. | *0.012* | *0.025* |
| | Slouched | 259.0 ± 79.4 | 263.9 ± 79.1 | 247.9 ± 73.3 | *0.001$^a$* | n.s. | n.s. | *0.008* |

B1 – first bump; B2 – second bump; B3 – third bump; a - Mauchly's Test of Sphericity; b - Greenhouse-Geisser test;

**Acknowledgement**: We acknowledge the support of Toyota Motor Corporation.